\documentclass[onecolumn,journal,12pt]{IEEEtran}

\usepackage{setspace}
\ifCLASSOPTIONonecolumn \doublespace \fi
\usepackage{floatflt}
\usepackage{wrapfig}
\usepackage{graphics}
\usepackage{rotating}
\usepackage{amsfonts}
\usepackage{amsmath}
\usepackage{subfigure}
\usepackage{comment}

\graphicspath{{Figures/}} \DeclareGraphicsExtensions{.eps,.ps}

\def\urltilde{\kern -.15em\lower .7ex\hbox{\~{}}\kern .04em}

\begin{document}
\title{Green Cellular Wireless Networks: Where to Begin?}
\author{\IEEEauthorblockN{Ender Ayanoglu}\\
\ifCLASSOPTIONonecolumn
\IEEEauthorblockA{Center for Pervasive Communications and Computing\\
Department of Electrical Engineering and Computer Science\\
University of California, Irvine
}
\else
\thanks{The author is with the Center for Pervasive Communications and Computing, Department of Electrical Engineering and Computer Science, University of California, Irvine, CA 92697-2625, USA.}
\fi
}
\maketitle
\begin{abstract}
%
\boldmath
Conventional cellular wireless networks were designed with the purpose of providing high throughput for the user and high capacity for the service provider, without any provisions of energy efficiency. As a result, these networks have an enormous Carbon footprint. In this note, we describe the sources of the inefficiencies in such networks. First we quantify how much Carbon footprint such networks generate. We also discuss how much more mobile traffic is expected to increase so that this Carbon footprint will even increase tremendously more. We then discuss specific sources of inefficiency and potential sources of improvement at the physical layer as well as higher layers of the communication protocol hierarchy. In particular, considering that most of the energy inefficiency in wireless cellular networks is at the base stations, we discuss multi-tier networks and point to the potential of exploiting mobility patterns in order to use base station energy judiciously.
\end{abstract}
\section{Introduction}\label{sec:intro}

The development of Information and Communication Technologies (ICT) within the last few decades has improved our lives tremendously, made information highly accessible, and increased productivity to unprecedented levels. It is expected that this trend will continue. However, this extraordinary improvement in our lives has a hidden cost. ICT employs computers, their peripherals, and communications equipment, all of which use energy, in many cases even when they are idle. As a result, energy consumption, and therefore the generation of greenhouse gases by this technology are already at very high levels. It is currently estimated that the ICT industry is responsible from about 2-4\% of all of the Carbon footprint generated by human activity \cite{W-GREEN}. This corresponds to about 25\% of all car emissions and is approximately equal to all airplane emissions in the world \cite{W-GREEN}. This trend will only increase. Internet equipment manufacturer Cisco Systems Inc. predicted in 2008 that the Internet traffic will increase annually 46\% from 2007 to 2012, nearly doubling every two years \cite{Cisco08}. Andrew Odlyzko of the University of Minnesota, who has been closely observing the growth of the Internet since the 1990s, has been quoted in 2008 as expecting an annual Internet traffic growth rate of 50-60\% \cite{Malik08}. With the proliferation of smart phones, video, and social networking, this rate of increase can be expected to be at least sustained for many years to come. As a result, serious concerns about the Carbon footprint impact of this development have been raised, and the topic of ``Green Communications" has been attracting attention in ICT circles (see, e.g., \cite{GreenTCSC,GreenConf}). Increasingly, more workshops, conferences, special issues of magazines and journals, and industry initiatives focus on the need to develop new approaches to communications and networking that result in drastically lower energy consumption. As the current communications and networking systems and protocols were not designed with this consideration, and since in many cases a greenfield approach needs to be taken, this effort will likely take a long time.

\begin{figure}[!t]
\begin{minipage}[b]{0.45\linewidth}
\centering
\includegraphics[height=45mm]{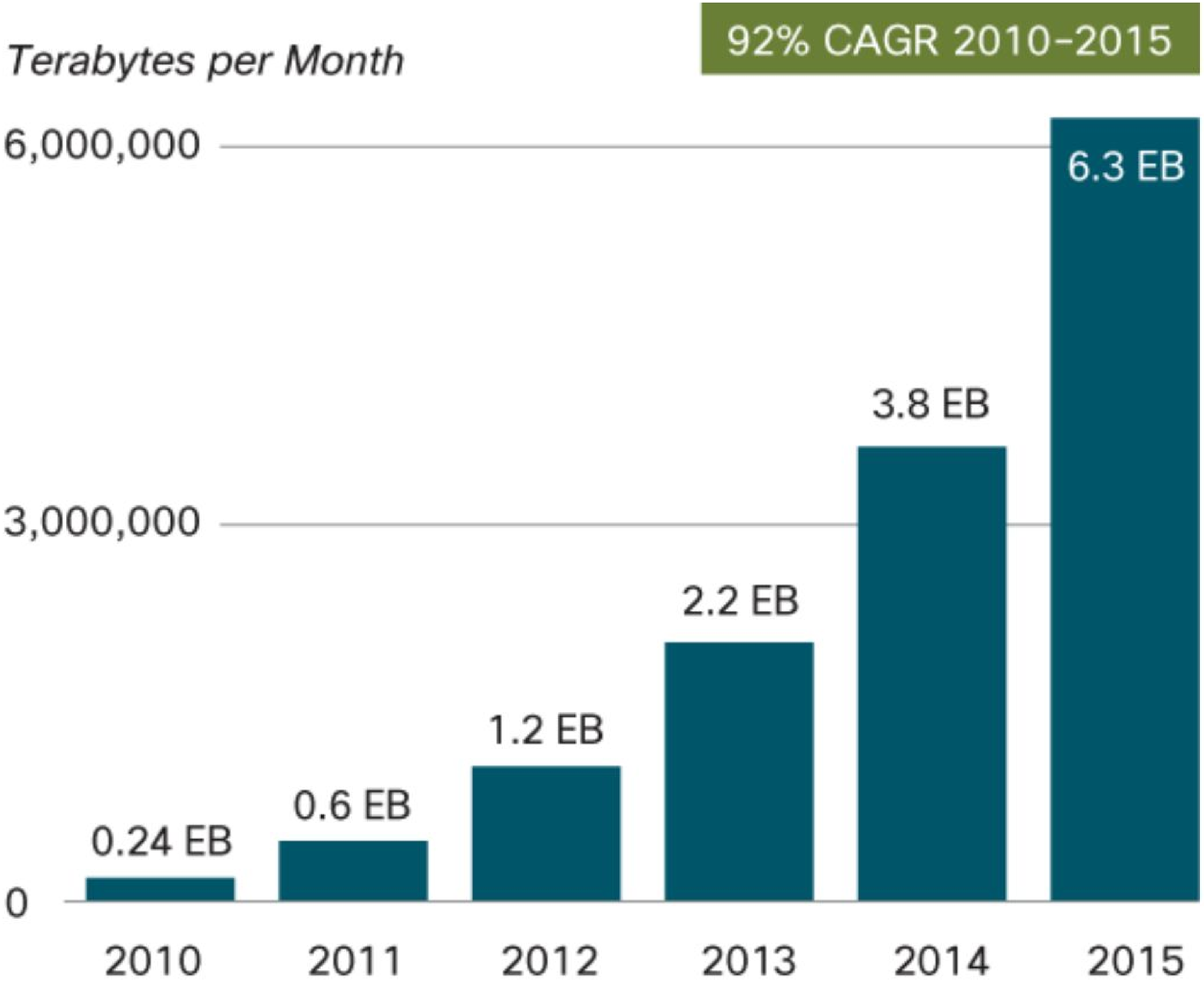}
\caption{Expected growth of wireless data \cite{Cisco11}.}
\label{fig:wirelessdata1}
\end{minipage}
\hspace{5mm}
\begin{minipage}[b]{0.45\linewidth}
\centering
\includegraphics[height=45mm]{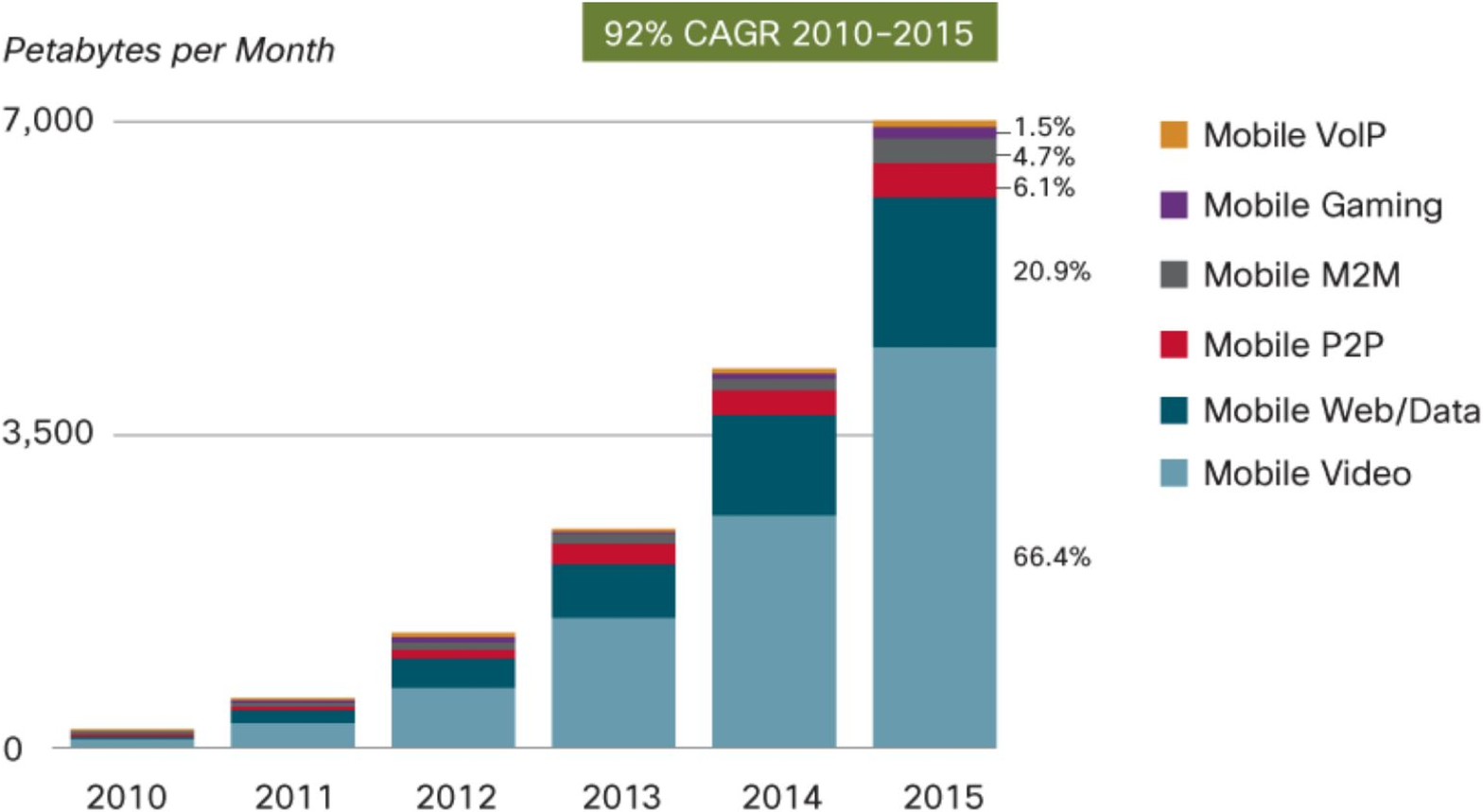}
\caption{Expected growth of mobile video \cite{Cisco11}. A Petabyte is $10^{15}$ bytes.}
\label{fig:wirelessdata2}
\end{minipage}
\end{figure}

In the case of wireless communications and networking, the increase in traffic is even more than the Internet as a whole. It is estimated that wireless data volume in the cellular segment alone is doubling annually. A study by Cisco Systems, Inc. states that the expected Compound Annual Growth Rate (CAGR) in the wireless service sector during 2010-2015 is 92\% \cite{Cisco11}. This is depicted in Fig.~\ref{fig:wirelessdata1}. In this figure, a Terabyte is $10^{12}$ bytes and an Exabyte (EB) is $10^{18}$ bytes. As a result, {\em wireless Internet traffic in 2015 will be 40 times that of 2010.\/} Although this growth is dramatic, even bigger growth rates have been observed in the past. For example, the United Kingdom-based service provider O2 reported that its mobile data traffic in Europe doubled every three months in 2009; Telecom Italia announced that its mobile traffic grew 216 percent from mid-2008 to mid-2009; and AT\&T has reported that its mobile traffic increased 5000 percent in the past 3 years (see, e.g., \cite{Schenker10}).

As detailed in Fig.~\ref{fig:wirelessdata2}, about 2/3 of the traffic in 2015 is expected to be due to mobile video \cite{Cisco11}. This is somewhat surprising since technologists questioned the value of mobile video in the past. However, the development of smart phones and the unexpected evolution of social networking have changed this picture in a major way. The average smart phone user generates 24 times the amount of traffic generated by the average non-smart phone user, with tablets this is 122 times \cite{Cisco11}. Handset traffic is highest in regions with the highest smart phone penetration. In addition to smart phones and tablets, e-readers, wireless video cameras, laptop computers, and a recent feature addition to smart phones, mobile phone projectors, are expected to be responsible from this increase in wireless traffic. This enormous increase in traffic will place a similarly enormous burden on the wireless service infrastructure. New approaches for coming up with new technologies will certainly be necessary to address these expected needs.

Conventional designs of mobile wireless networks mainly focused on ubiquitous access and large capacity, or high throughput to the user, without any considerations of power or energy efficiency. But, with the inevitable constant use of ICT in our lives, there needs to be a serious thought given to increasing the energy efficiency in these technologies. All major ICT manufacturers realize the importance of this problem and are planning to take steps in this direction. Examples are Cisco Systems, Inc. \cite{CiscoGreen}, Ericsson \cite{EricssonGreen}, Huawei \cite{HuaweiGreen}. Alcatel-Lucent is leading a consortium with partners from industry and academia to generate a plan to reduce power consumption in ICT drastically, by three orders of magnitude, by the year 2015 \cite{GreenTouch}. Alcatel-Lucent states this is based on an internal study, based on Shannon theory, which showed four orders of magnitude reduction can be realized.

It may be possible to come up with a number of simple protocol changes and introduce sleep cycles into existing protocols. An important example along those lines is the IEEE 802.3 Ethernet protocol. This protocol is commonly used to network personal computers, servers, and various peripherals at speeds of 10 or 100 Mb/s, or 1 or 10 Gb/s. In its original version, it was designed to transmit physical layer signals even when there is no information to be transferred. With the highly bursty nature of data communications, this is unnecessary and wasteful. The IEEE 802.3az version of the standard changed that by sending a Low-Power-Idle (LPI) indication signal for a specified time and consequently allowing the transmit chips in the system to be turned off. LPI is sent periodically to refresh the sleep mode. When there is data to transmit, a normal idle signal is sent to wake the transmit system up before data is due to be sent. The data link is considered to be always operational, as the receive signal circuit remains active even when the transmit path is in sleep mode. The standard was approved as IEEE Standard 802.3az-2010 at the September 2010 IEEE Standards Board meeting \cite{802.3az} and products are available from various manufacturers. The savings will be small at the onset due to the presence of legacy equipment but they are expected to rise to \$410M/year for the U.S. and over \$1B/year worldwide, exceeding the cost of current energy used by Ethernet physical layer components \cite{Christensen10}. In an attempt to introduce power savings for 802.11 Wi-Fi Wireless Local Area Networks (LANs), reference \cite{Haratcherev09} notices that most Wi-Fi Access Points (AP) are always on. It also makes the estimation that an active daily session for a home AP is about 91 minutes. It introduces a low-power sleep mode concept and shows through a prototype that average power could be reduced from 3.36 Watts to 2.48 Watts between the interface being on and off. A lot more power reductions are possible when there are extended periods of no use, e.g., when the residents are not home or when they are sleeping. Reference \cite{Haratcherev09} recognizes this and targets a 90\% power reduction as future work.

In this proposal, we attack the problem of significant power or energy reduction in cellular wireless networks. We first identify and discuss sources of potential and significant energy reduction in such networks. We present a new access technique and a different network topology to eliminate existing major power losses. We discuss a dynamic network architecture that will adapt to use patterns in order to reduce energy consumption. The dynamic nature of this architecture allows it to sense the environment and adapt. We present a physical layer technique to operate power amplifiers efficiently.
We also discuss directional antenna techniques for significant energy reduction in power amplifiers of Radio Base Stations (RBS) and Remote Terminals (RT). 
\section{Sources of Energy Inefficiency in Cellular Networks}\label{sec:losssources}

\begin{figure}[!t]
\begin{minipage}[b]{0.3\linewidth}
\centering
\includegraphics[height=33mm]{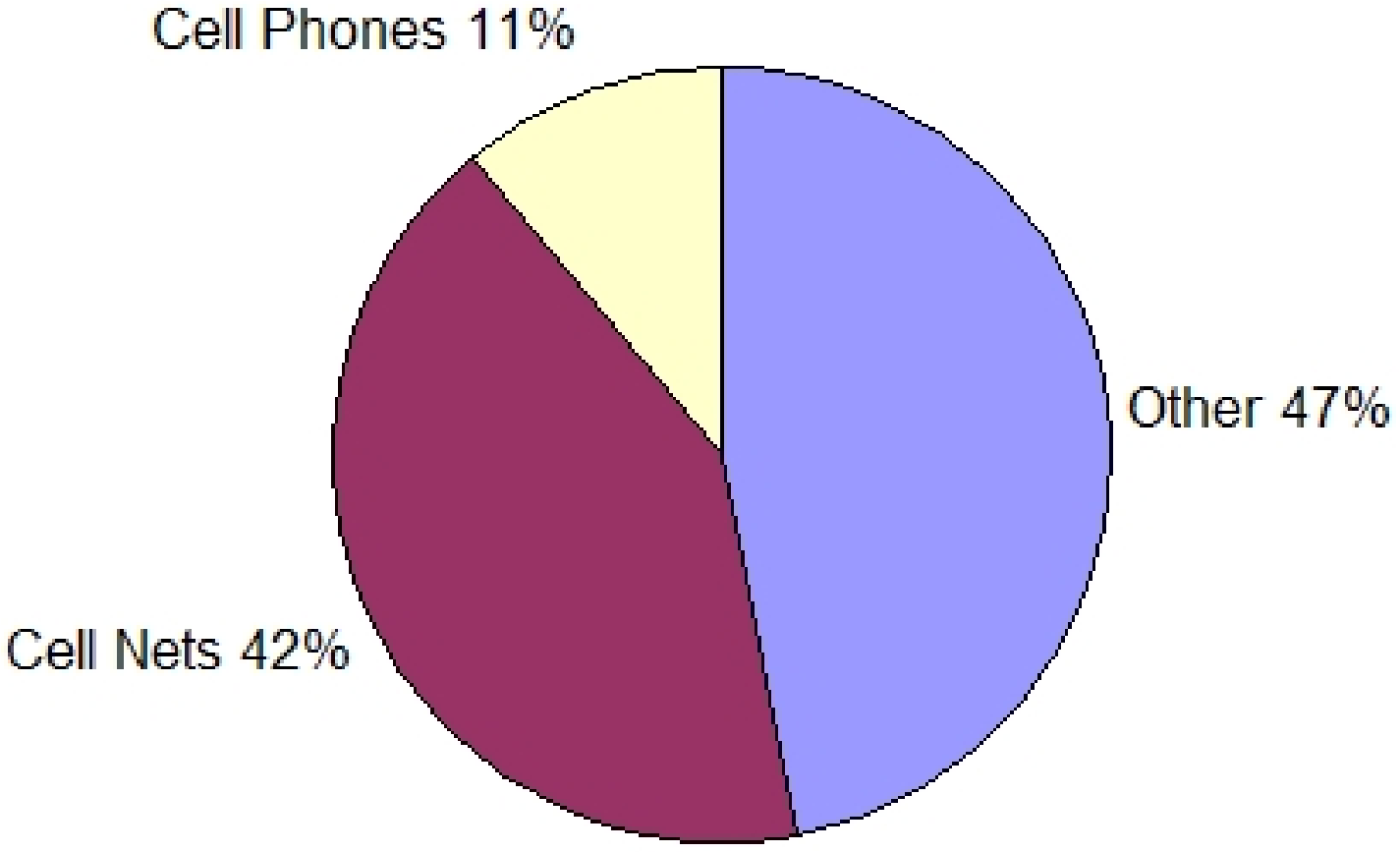}
\vspace{1mm}
\caption{Percentage of MtCO$_2$e contributions in 2002.}
\label{fig:MtCO2e02}
\end{minipage}
\hspace{1.6mm}
\begin{minipage}[b]{0.3\linewidth}
\centering
\includegraphics[height=33mm]{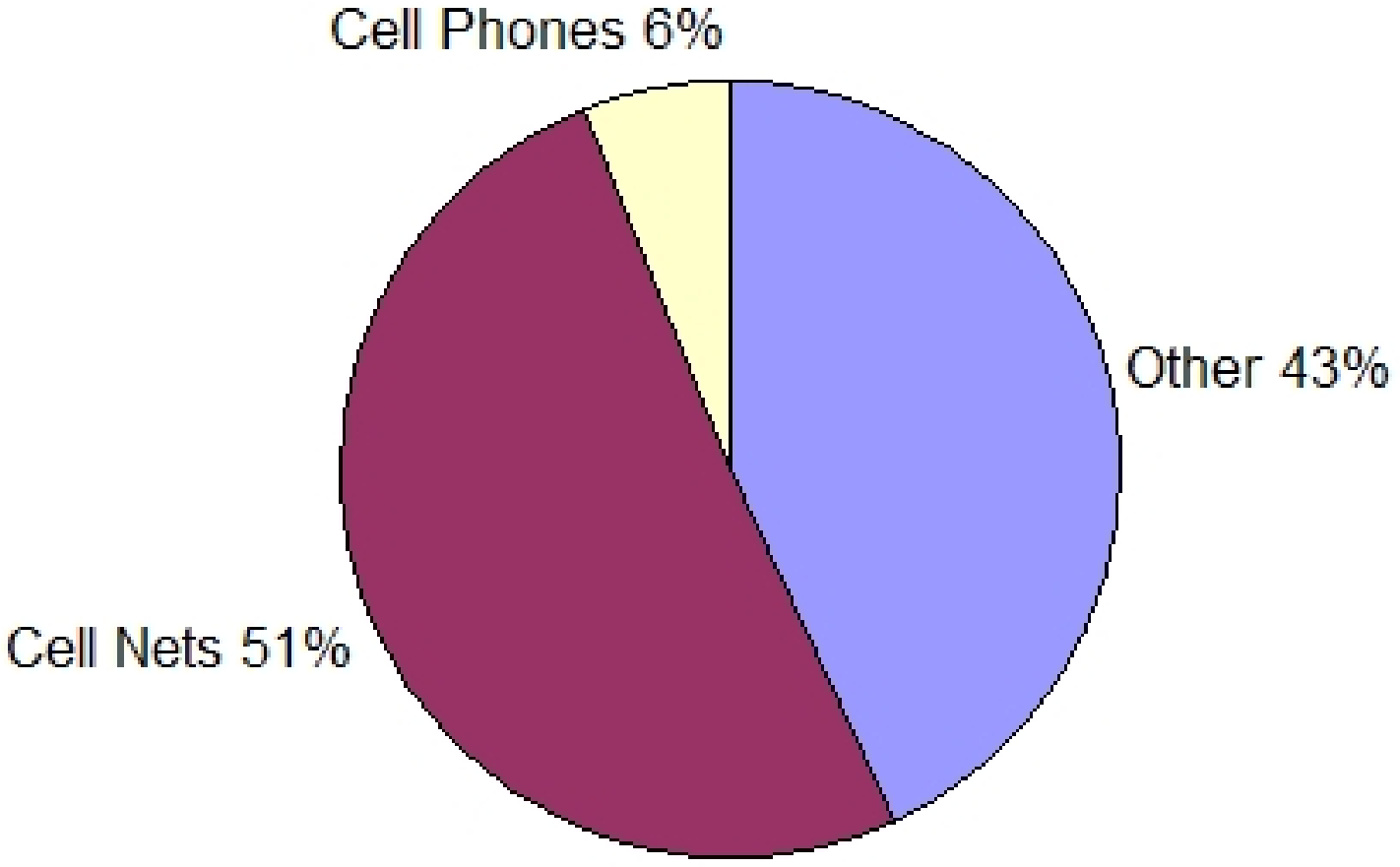}
\vspace{0.2mm}
\caption{Percentage of MtCO$_2$e contributions in 2020.}
\label{fig:MtCO2e20}
\end{minipage}
\hspace{5mm}
\begin{minipage}[b]{0.3\linewidth}
\centering
\mbox{\hspace{-3mm}}\includegraphics[height=40mm]{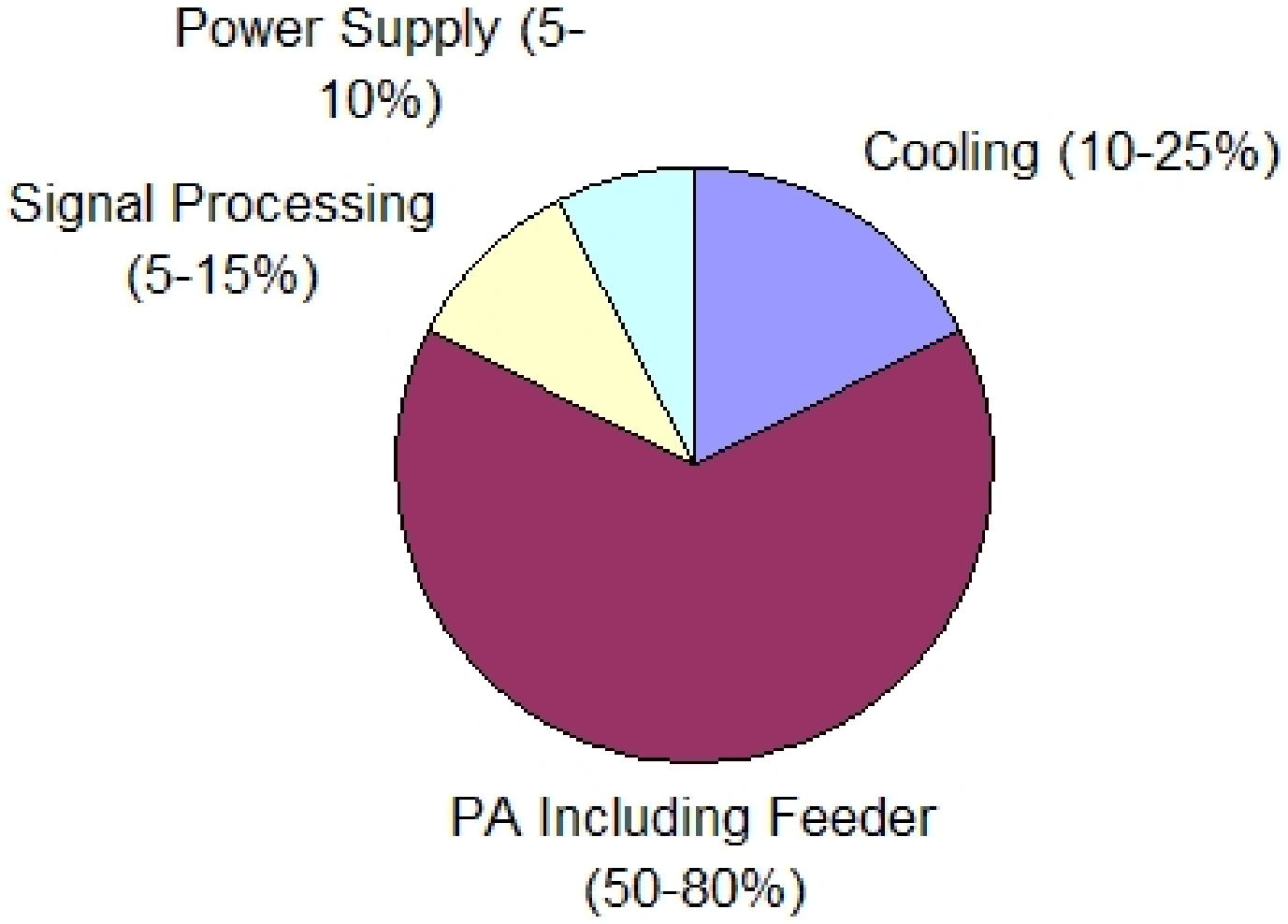}\\
\vspace{-3mm}
\caption{Distribution of power consumption in a cellular base station.}
\label{fig:PAPower}
\end{minipage}
\end{figure}

In 2002, telecommunications infrastructure and devices (not including data serves, PCs, and peripherals) were responsible from 151 Metric tonne (ton) CO$_2$ equivalent (MtCO$_2$e) emissions, with cellular networks contributing 64 Mt (42\%) and mobile phones contributing 16 Mt (11\%), as depicted in Fig.~\ref{fig:MtCO2e02}. In 2020, the telecommunications infrastructure and services contribution will more than double and reach 349 MtCO$_2$e with mobile cellular networks contributing 179 Mt (51\%) and mobile phones 22 Mt (6\%) as depicted in Fig.~\ref{fig:MtCO2e20} \cite{Barth09}. This shows three things. First, the contribution due to cellular networks and mobile phones will more than double by 2020 in absolute terms. Second, this contribution is actually increasing in relative terms. And third, the contribution of cellular networks is far greater than mobile phones. Reference \cite{Barth09} states 80\% of the total power in cellular networks is consumed at base stations. Furthermore, \cite{Barth09} states that, in a cellular base station, the Power Amplifier (PA), together with its feeder, consumes 50-80\% of the total power, the remainder going to cooling (10-25\%), signal processing (5-15\%), and power supply (5-10\%), as shown in Fig.~\ref{fig:PAPower}. The distributions quoted by a number of other sources are similar (see, e.g., \cite{Grant09}). As a result of the discussion and the quoted figures above, it makes sense to {\em target energy efficiency in cellular base stations and particularly for issues related to the power amplifier, as the highest priority\/} in this proposed research. Conventionally, in cellular networks, base stations are designed with the goal of maximizing capacity or user throughput, and reducing capital expenses and operating expenses (commonly abbreviated as CapEx and OpEx, respectively). However, because of the increasing amount of traffic, OpEx are constantly increasing. The critical point is that increasing the user throughput makes conventional base stations operate in an inefficient mode. This is a problem that needs to be solved. To make matters a lot worse, newer generations of cellular wireless technologies are promising increasingly higher transmission speeds at higher distances from the RBS. In order to satisfy these needs, RBS PAs need to be operated in a linear mode where they are highly inefficient and consume a lot more power. As a result, it is quite clear that there is a need for a major change in the design of base RBSs of the future.

Currently available cellular networking technologies are known as 2nd or 3rd Generation, or 2G and 3G. There are two major standards supported, one is known as Code Division Multiple Access (CDMA). The other is based on Time Division Multiple Access and is known by the name Global System for Mobile Communications (GSM). In its 3G version, GSM adopted techniques from CDMA and the resulting set of techniques is given the name Wideband CDMA (WCDMA). GSM, CDMA, and WCDMA offer relatively low transmission rates of about 1-2 Mb/s per user by employing modulation alphabets with small constellation sizes. As a result, the burden on the base station for transmission with them is not as significant as the new technologies that are expected to replace them within a few years. Known as 4th Generation (4G), two new standards have emerged to replace 2G and 3G. Both of these standards have very aggressive goals for transmission rates to be offered to users. The first of these is developed around the IEEE 802.16 standard and is known as Worldwide Interoperability for Microwave Access (WiMAX). The current version of WiMAX can provide up to 40 Mb/s to the user. The next version, to be available when the IEEE 802.16m standard is completed, is poised to offer up to 1 Gb/s fixed (not mobile) access. The second standard, developed through an industry consortium, is known as Long Term Evolution (LTE). The term Long Term Evolution is intended to mean the evolution of the GSM technology, which has the largest user base in the world. LTE is designed to provide up to 50 Mb/s on the uplink and 100 Mb/s on the downlink. Its advanced version LTE-Advanced is planned to provide up to 1 Gb/s fixed access. In order to provide these kinds of transmission rates, both WiMAX and LTE employ sophisticated techniques with modulation schemes employing large constellation sizes and
a broadband wireless technology known as Orthogonal Frequency Division Multiplexing (OFDM). OFDM removes the need for long training and equalization periods to overcome the limitations of a frequency selective channel and is therefore without competition when broadband wireless services are to be offered. However, {\em the current direction of offering OFDM via legacy base stations covering a range of 3-8 mile radius and with aggressive transmission rates of 10s or even 100s of Mb/s to the users is questionable, especially from an energy consumption viewpoint.\/} There is no doubt that this push will enormously increase energy consumption in cellular wireless systems, which is already at alarmingly high levels, as discussed earlier.

\subsection{The Impact of Modulation on Power Amplifiers}
OFDM transmits signals over multiple subcarriers simultaneously. The advantage this provides can be thought of as dividing a broadband spectrum into smaller bands, each with a lot less frequency response variation. As a result, the requirement of equalizing the broadband frequency spectrum is substantially simplified. However, the multi-carrier nature of OFDM also has drawbacks. One of these drawbacks is the increased Peak-to-Average Power Ratio (PAPR). A peak in the signal occurs when all, or most, of the subcarriers are aligned in phase. The worst-case value of the PAPR is directly proportional to the number of carriers and is given by $\textrm{PAPR}=10 \log N$ where $N$ is the number of subcarriers \cite{Wight01,Tellambura}. For example, in the commonly used IEEE 802.16g Wi-Fi wireless LAN standard, there are $N=64$ subcarriers. In reality only 52 are used for data transmission, but for this argument and for simplicity we can assume all $N=64$ are. This means a worst-case PAPR of about 18 dB results. The PA must be able to accommodate this peak power as well as lower power levels, at least by an amount equal to PAPR. Typically, the operating point of the PA is set at the average level and the peak point is set at a level higher by an amount corresponding to PAPR in dB. This is the power level the PA loses its linear behavior. This is called power backoff. Backing off the power in this manner causes PAs to be employed for the transmission of OFDM signals to be more expensive since they should be able to accommodate a wider power range in their linear operating regions. On the other hand, linear power amplification is highly power inefficient. As a result, power consumption of the PA increases substantially, resulting in poor power efficiency of the overall system.

\begin{figure}[!t]
\begin{center}
\includegraphics[width=60mm]{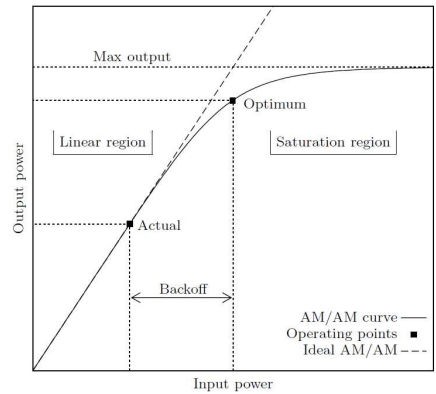}
\caption{Power amplifier transfer function and power backoff.}
\label{fig:Backoff}
\end{center}
\end{figure}

A representation of backoff is depicted in Fig.~\ref{fig:Backoff} \cite{Thompson05}. Ideally the output of the PA is equal to the input times a gain factor. In reality, the PA has a limited linear region, beyond which it saturates to a maximum output power level. Fig.~\ref{fig:Backoff} shows a representative input/output curve, known as the AM/AM conversion. The most efficient operating point is actually the saturation point of the PA, but for signals with large PAPR, the operating point must shift to the lower power region on the left, keeping the amplification linear. The average output power is reduced by an amount equal the power backoff value. To keep the peak power of the input signal less than or equal to the input saturation level, the power backoff must at least be equal to the PAPR. As discussed in \cite{Thompson05}, for an OFDM system with $N=16$, the PAPR and therefore the required power backoff is about 9.5 dB, excluding the subcarriers that do not carry data (or about 12 dB including all of $N=16$ subcarriers). At this point, {\em the efficiency of a Class A PA is less than 6\%. This is going to get much worse when the number of subcarriers increases.\/} We would like to note that both WiMAX and LTE have incorporated larger numbers of subcarriers, up to $N=2048$. This increases PAPR to approximately 33 dB. Although the worst case PAPR is uncommon and practical PAPR values are several dBs below the worst-case value of $10\log N$, the change in practical PAPR values are still linear with the number of subcarriers $N$ \cite{Goldsmith}, and therefore the increase in the number of subcarriers from 64 to 2048 will have a substantial negative impact on power efficiency of OFDM PAs, especially at the base stations.

In addition, PAPR also increases slightly with increasing the constellation size \cite{Tellambura}. However, increasing the constellation size increases the power consumption in a PA. For example, it is well-known that for bandlimited channels, for modulation techniques such as Quadrature Amplitude Modulation (QAM), and for the same Bit Error Rate (BER), as the constellation size increases, a larger value of bit energy is needed \cite{Proakis}. Therefore, increasing the constellation size requires an increased power level at the receiver. The received power in a wireless system is proportional to the transmitted power, based on the distance between the transmitter and the receiver as well as a descriptor of the channel known as path loss. As a result, increasing the constellation size increases the power consumption, and therefore the power inefficiency, at the transmitter. In most cases, the transmission rate from the RBS to the RT will be higher. Furthermore, as discussed earlier, the power inefficiency at the RBS is the more significant problem and we will concentrate on that problem in the proposed research. QAM is employed in each subcarrier of an OFDM system and therefore the conclusion that increasing the constellation size increases the power inefficiency holds for OFDM. The OFDM techniques used in WiMAX and LTE employ channel coding. With channel coding, the conclusion that an increase in constellation size increases the power inefficiency holds. In order to see this, we refer the reader to simulation results of a system similar to that used in WiMAX as well as LTE, including channel coding, in \cite{Ayanoglu01}, or in \cite{Ayanoglu00} with clearer plots, based on the PI's past work.

Communication theory indicates that in the case of power-limited channels, modulation techniques optimum in the case of bandlimited channels, such as QAM or Phase Shift Keying (PSK) are actually not optimal. In this case, the optimum modulation technique is known as orthogonal signaling where each message is assigned a distinct signal and all of these signals are orthogonal to each other \cite{Proakis}. Although the word orthogonal appears in the expansion of the acronym OFDM, OFDM is not orthogonal in this particular sense because each of its orthogonal signals carry more than one message in general. An example of achieving this orthogonality is Frequency Shift Keying (FSK) where the signals are sinusoids with carrier frequencies spaced apart so that they are orthogonal over a symbol period. One can employ $M$ orthogonal signals to generate a modulation technique known as MFSK. The resulting signal will have continuous amplitude (or envelope) and it is desirable to make such signals to have continuous phase as it leads to better performance when nonlinear, efficient PAs are used to transmit them. The tradeoff for this power advantage is the increased bandwidth due to orthogonality.  In wireless sensor networks where transmission distances are very short, of the order of meters, and transmission rates are very low, of the order of bits per second, interesting tradeoffs exist that can favor MFSK together with transmission time tradeoffs and with or without channel coding (see, e.g., \cite{CGB05}). These techniques can indeed be applicable to wireless sensor networks due to short distance transmission and therefore the possibility of reusing the frequency spectrum over space as well as low bit rates. However, for cellular transmission, the use of MFSK would be inefficient, even prohibitive, because of the tradeoff of bandwidth. However, a constant envelope version of OFDM could be an interesting choice. There are recent developments in a constant amplitude version of OFDM and we will discuss this approach later in this proposal.

\subsection{Deployment, Frequency Plan, Mobility, and Traffic Patterns}
Reference \cite{CZXL11} states a fundamental tradeoff in realizing energy-efficient networks is a tradeoff between energy and deployment. We agree that this is a very important issue. One of the most significant inefficiencies in cellular networks today has to do with RBS deployment. Therefore, we will next summarize the way cellular networks are deployed (see, e.g., \cite{Rappaport}).

\begin{figure}[!t]
\begin{center}
\begin{minipage}[b]{0.3\linewidth}
\centering
\mbox{\hspace{1mm}}\includegraphics[width=36mm]{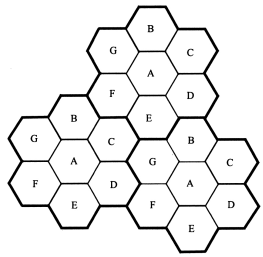}
\caption{Frequency reuse.}
\label{fig:reuse}
\end{minipage}
\end{center}
\end{figure}

\begin{figure}[!t]
\begin{center}
\begin{minipage}[b]{0.3\linewidth}
\centering
\mbox{\hspace{1mm}}\includegraphics[width=36mm]{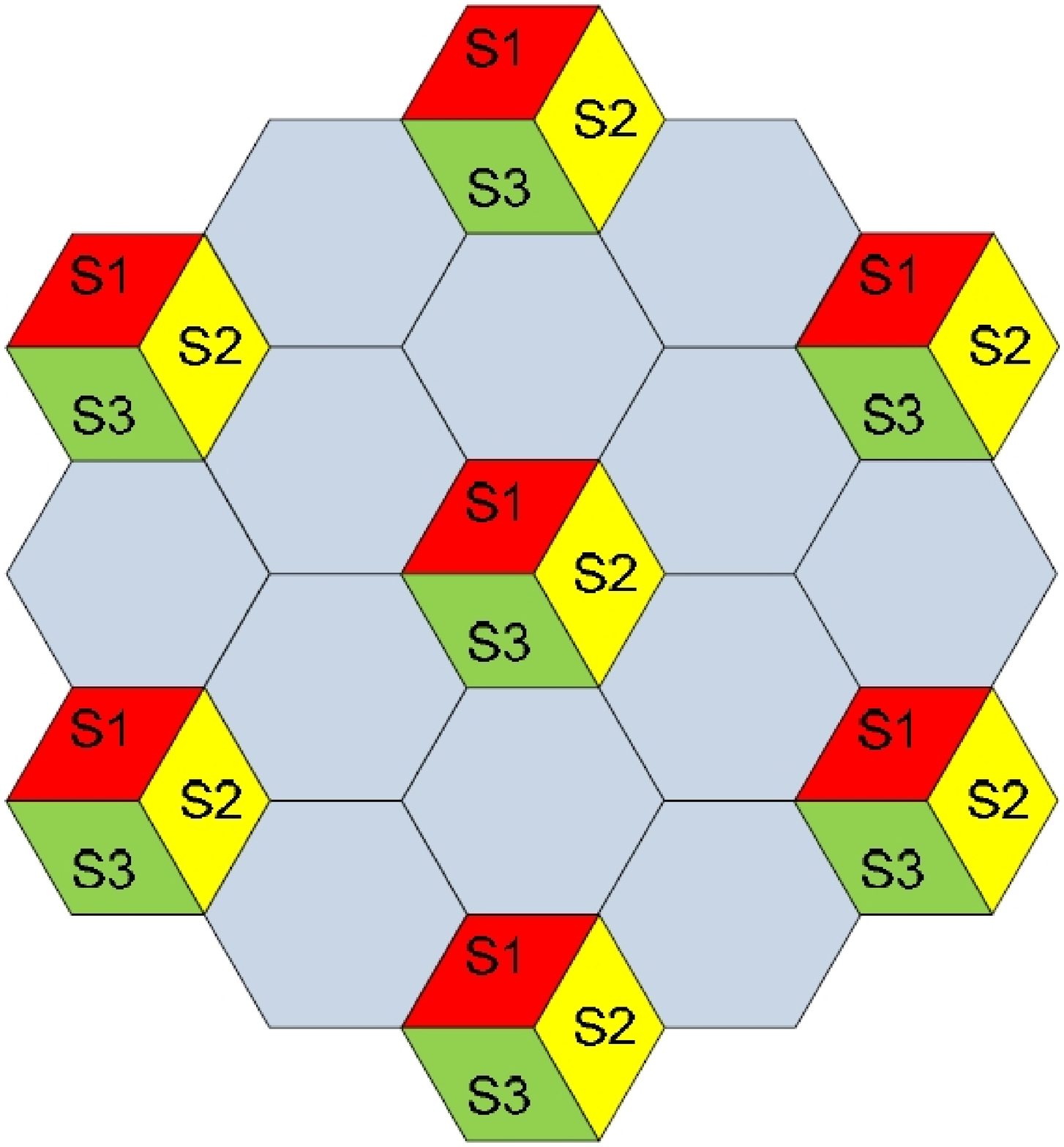}
\caption{Sectoring.}
\label{fig:sector}
\end{minipage}
\hspace{0.5cm}
\begin{minipage}[b]{0.3\linewidth}
\centering
\includegraphics[width=60mm]{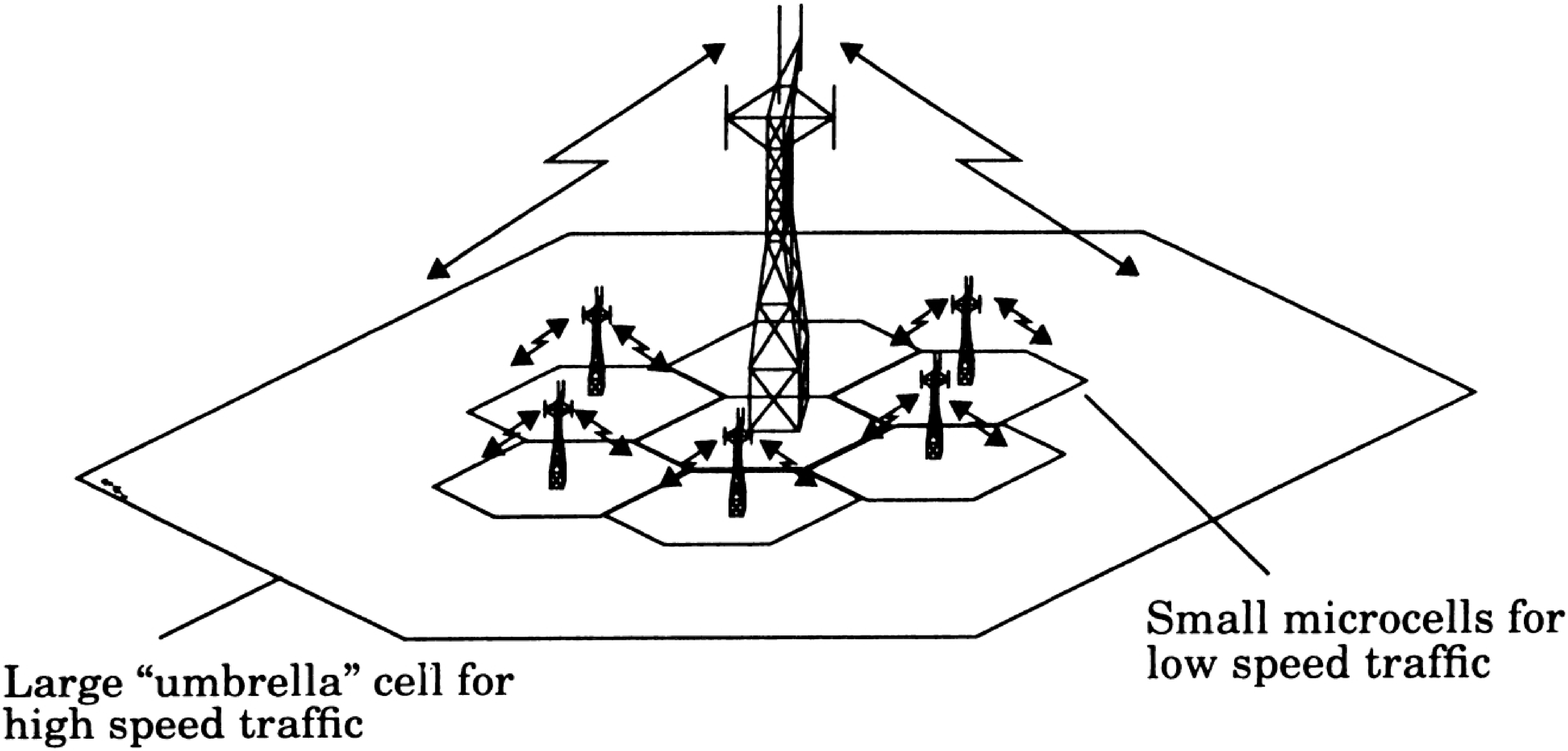}
\caption{Umbrella network.}
\label{fig:umbrella}
\end{minipage}
\end{center}
\end{figure}

Early mobile telephony systems consisted of a single, high powered transmitter with an antenna mounted on a tall tower. This provided wide coverage but could only serve a small number of channels because it did not allow for the reuse of the spectrum at different physical locations in a wide area. The concept of a cellular network was introduced to solve this problem. This concept replaces a single, high power transmitter (which can be interpreted as a large cell) with many low power transmitters (or small cells), each providing coverage to a small part of the service area. Every cell has an RBS at its center. Each RBS is allocated a portion of the total number of channels so that all the available channels are assigned to a small number of neighboring RBSs. Neighbor RBSs are assigned different groups of channels so that the interference between RBSs (and RTs in the area they control) is minimized. As a result, the total available number of channels are reused throughout a coverage region. Each RBS is allocated a group of channels to be used within a small geographic area, technically called a {\em cell.\/} Fig.~\ref{fig:reuse} illustrates the concept of frequency reuse. Each of the hexagonal cells labeled from A through E are allocated a set of different frequency channels. The distances are chosen such that the interference among cells with the same frequency designation can be kept to a minimum. Today's cellular networks employ cell sizes with about 3-8 mile radius. The hexagonal cell shapes are conceptual and in reality frequency planning, taking the effects of the terrain and shadowing due to buildings, is carried out. In order to reduce interference among cells further, a technique known as
{\em sectoring\/} is employed. In this technique, each cell is further subdivided into sectors, as shown in Fig.~\ref{fig:sector} with 120 degrees. Each RBS employs a directional antenna transmitting and receiving only within the sector angle. As a result, interference among RBSs is reduced. For example, in the 120$^\circ$ configuration shown in Fig.~\ref{fig:sector}, sectoring reduces the number of interfering RBSs from 6 to 2.

At this point, we would like to emphasize that the ideas of cellular networks and frequency reuse are solid as far as efficient energy consumption is concerned. The problem, especially looking into the future, lies with the cell radii, available PA technology, intended transmission rates, and user mobility. Although transmission rates demanded by users in a fixed location or moving at pedestrian speeds may be large, transmission rates demanded by users moving at vehicular speeds will always be limited. Users who are in public transport vehicles such as trains, buses, or planes can be served by in-vehicle APs with high-rate connections to the backbone via wireline or, for example, satellite connections. Users driving cars cannot demand transmission rates as high as multiple Mb/s or higher for safety concerns. As a result, trying to serve a limited number of users in automobiles demanding very high rates using the cellular network can be avoided. Such users can also be served via satellite, removing a big burden on the cellular network. As an example, we would like to quote the two-tier network, distinct for highly mobile users (in macrocells) and less mobile users (in micro or picocells), as in Fig.~\ref{fig:umbrella} \cite{Rappaport}. {\em Although this idea is appreciated by the cellular research and development community, it has actually not been fully developed or implemented.\/}

\begin{figure}[!t]
\begin{minipage}[b]{0.45\linewidth}
\centering
\includegraphics[width=60mm]{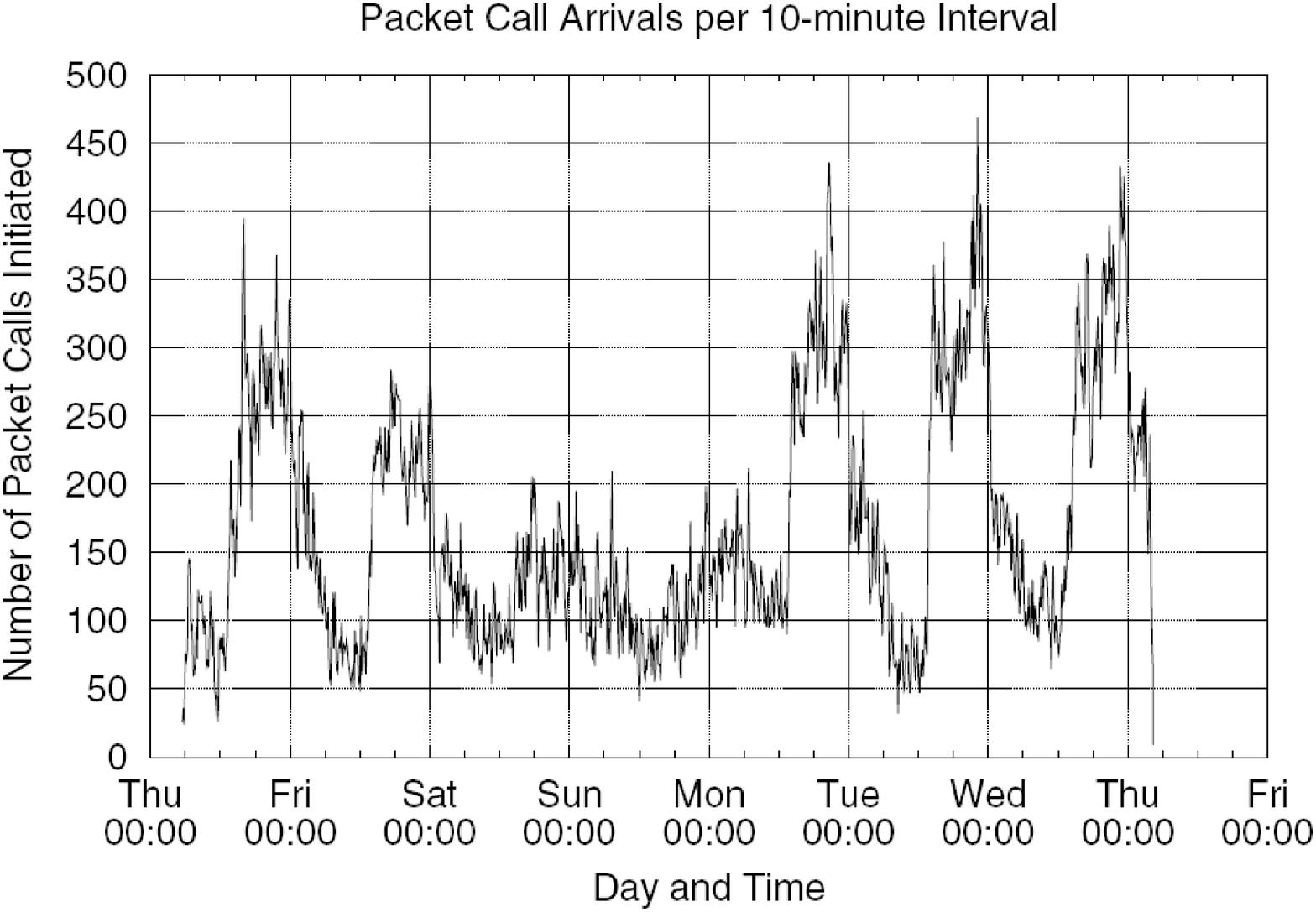}
\caption{Weekly traffic pattern at a cellular base station.}
\label{fig:traffic}
\end{minipage}
\hspace{5mm}
\begin{minipage}[b]{0.45\linewidth}
\hspace{10mm}\includegraphics[width=60mm]{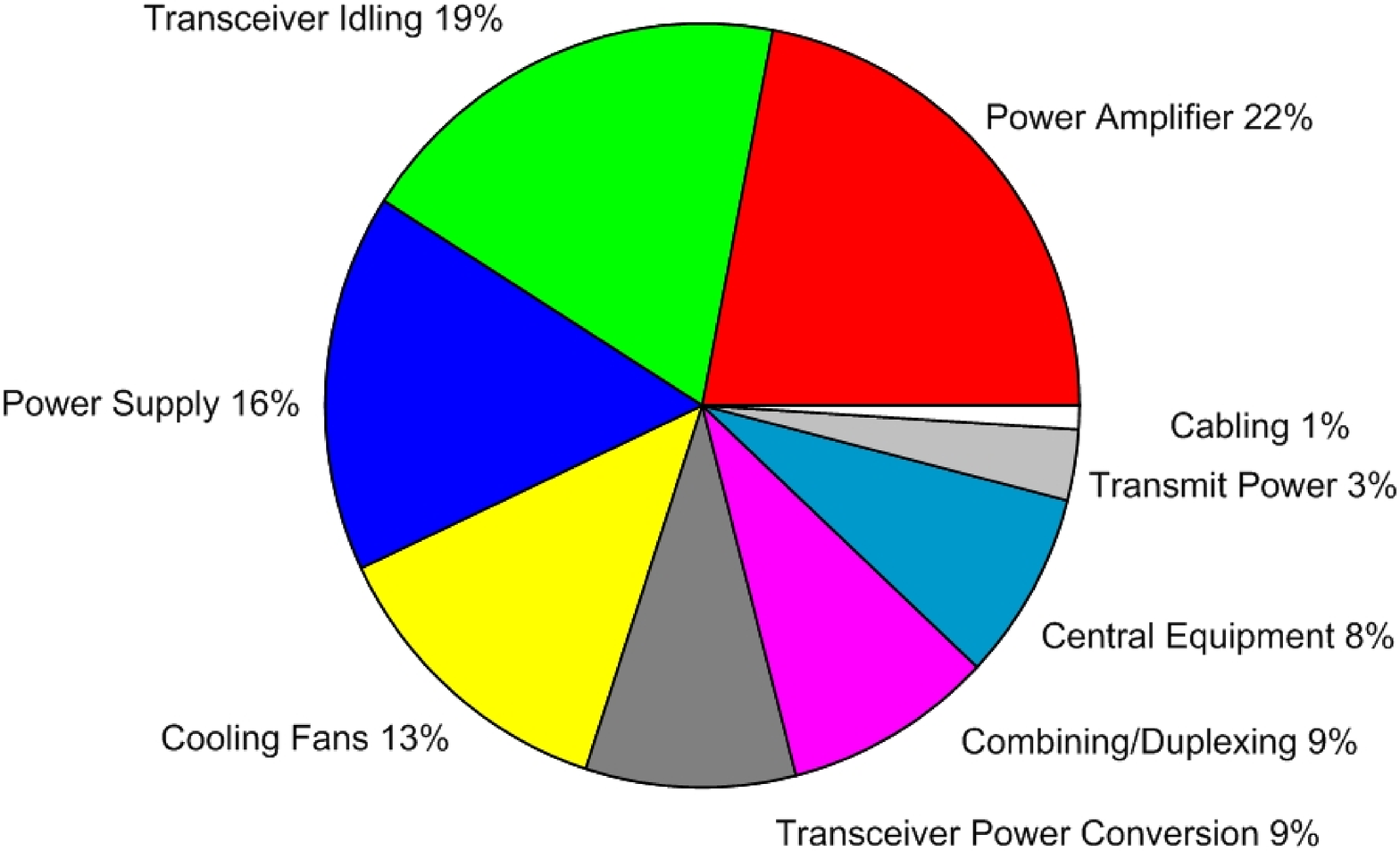}
\caption{Detailed distribution of power consumption in a cellular base station.}
\label{fig:grant}
\end{minipage}
\end{figure}

There is some data available on cellular network traffic, e.g., \cite{WHSW05,TRKP10,Correia10,OKLN11}, \cite[p. 250]{HT09}. As can be observed in Fig.~\ref{fig:traffic}, cellular network traffic shows a great deal of variation during the day and also, according to the day of the week \cite{WHSW05}. The traffic is time-varying, with heavier traffic occurring during weekdays as compared to weekends. Overall, there are significant periods of low network utilization. Heaviest traffic occurs during late afternoon and evening hours. This is different than the wired Internet, where traffic more closely aligns with normal working hours. The ratio of the heaviest traffic volume to the lightest is of the order of about 2 in \cite{WHSW05,HT09}, and significantly higher (about 6) in \cite{TRKP10}. {\em Currently, this is not exploited, and base stations remain idle during these periods.} In idle periods, RBSs keep consuming substantial power. A more detailed version of Fig.~\ref{fig:PAPower} is given in Fig.~\ref{fig:grant} \cite{Grant09}. This figure shows that 19\% of the power consumed is wasted in transmitter idling. This is potentially a substantial source of improvement.

In addition, there is traffic variation with respect to location as well. As a result, there is a large savings potential, and not only for periods of low traffic. According to an Alcatel-Lucent study, typically 10\% of the sites carry 50\% of all traffic;
in addition, 50\% of sites are lightly loaded, carrying only 5\% of the traffic \cite{Barth09}. {\em Currently, this difference is also not exploited.}

\subsection{RBS Energy Consumption and Its Effects for the Service Provider}
As described earlier, the cellular RBS is the most energy consuming component in a cellular network. A typical 3G RBS is highly inefficient. {\em For about 40 W output power, it consumes about 500 W of power, corresponding to only 8\% efficiency\/} \cite{Grant09}. As a result, the annual energy consumption of an RBS is around 4.4 MWh. It has been reported that there were approximately 52,500 base station sites at the end of 2009 in the United Kingdom \cite{MOA}. This results in an energy consumption of 230 GWh per year by the U.K. cellular network, corresponding to 165K MtCO$_2$e, or the CO$_2$ emissions equivalent of 31.5K cars \cite{epa}.
The numbers are larger for countries with larger populations. For example, in China, there are 500K GSM/EDGE and 200K 3G CDMA RBSs \cite{Grant09}. This results in 3.1 TWh, or 2.2M MtCO$_2$e, or the CO$_2$ emission equivalent of 421K cars. The number for the US is similar, about 750K \cite{TRKP10}. In addition to CO$_2$ emissions, energy consumption increases the system OpEx for service providers. Service providers will clearly want to improve the efficiency figure of 8\% for 3G base stations. As we discussed earlier, this figure is actually likely to go further down with 4G base stations.

\begin{figure}[!t]
\begin{center}
\includegraphics[width=60mm]{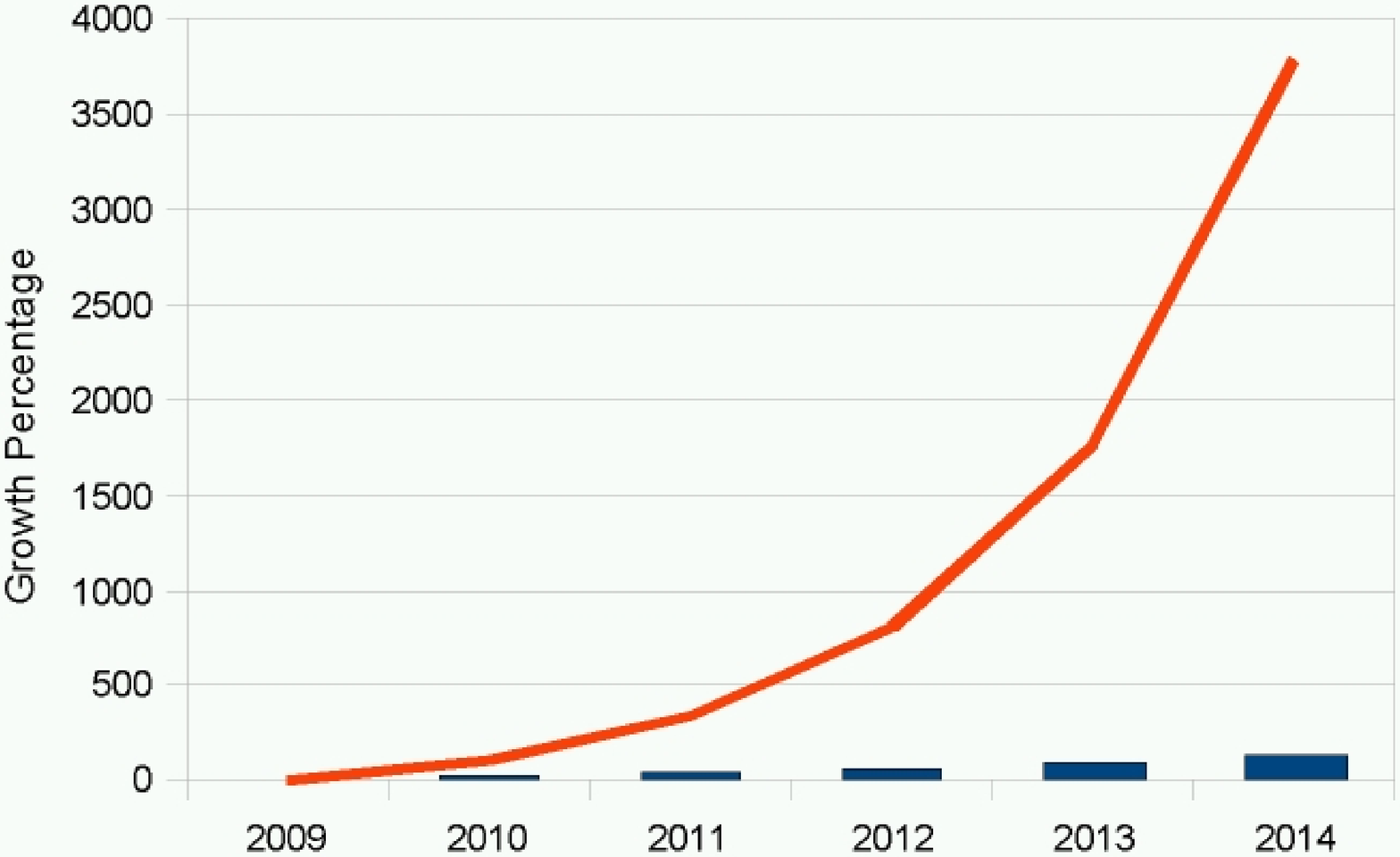}
\caption{Projected data traffic (red line) and revenue growth (blue bars) 2009-2014.}
\label{fig:cagr}
\end{center}
\end{figure}

We stated earlier that the expected growth rate for wireless data volume is 92\%. The projections for U.S. wireless data revenues for service providers are much less, typically around 10-18\% CAGR \cite{Smith10,SNL,ABI}. Similar expectations are made for other countries \cite{Grant10}. These projections result in an increase of 39 times for wireless data traffic in 2014 as compared to that of 2009 \cite{Cisco10}, whereas only 2.28 times for data revenue for the same period, as shown in Fig.~\ref{fig:cagr}. Wireless voice CAGR values are actually being projected as becoming negative \cite{Smith10}. As a result, reducing OpEx values is imperative for service providers. Since the same service providers will have to accommodate a tremendous increase in traffic, managing OpEx becomes very important. It can be argued that projections of data revenue CAGR were made without insights into explosion in data traffic. Even though this argument is somewhat valid, the fact remains that wireless service providers are under heavy competitive pressure to keep subscription fees low while at the same time are facing exponentially increasing traffic. So, reduction of service provider OpEx will certainly be extremely important.
\section{Conclusions}
ICT industry generates 2-4\% of all of the Carbon footprint generated by human activity, equal to about 25\% of all car emissions and approximately equal to all airplane emissions. With the increased use of the Internet, this trend is going to increase. For wireless cellular networks, the expected increase in traffic is even more than the Internet as a whole, doubling approximately every year. Due to the proliferation of smart phones, tablets, social networks, and mobile video, this trend is expected to hold for a long time.

Designs of cellular wireless networks were based on large user throughput and high service provider capacity, without any considerations for power or energy efficiency. In order to attack this problem, one needs to understand the sources of energy efficiency in these networks. To that end, it is known that most of the energy inefficiency is in the base stations. In base stations, the power amplifier is a large source of inefficiency. Since all of the wireless networks of the future will employ Orthogonal Frequency Division Multiplexing (OFDM), considerations of efficiency for the power amplifier using this modulation technique are important. It is well-known that OFDM results in large Peak-to-Average Power Ratio (PAPR). Although solutions were proposed for this problem, they have not gained widespread availability and a new and efficient technique can be very useful.

A major source of inefficiency in cellular wireless networks is the large cell radii of 3-8 miles. This structure, although originally designed for mobile users and telephony, is increasingly being used for users in fixed locations but with large traffic demands. A two-tier network where mobile users with vehicular speeds are supported by means of a large cell and users in fixed locations or at pedestrian speeds by means of smaller underlay cells can be solution for this problem. Algorithms for the operation and handoffs should be designed for such networks. Another major source of energy inefficiency that can be exploited is due to daily and weekly traffic patterns. These variations are major, showing major differences between say, day and night. It can be speculated that a number of algorithms, such as turning off base stations with low usage and handing off their users to base stations with large radii, can be devised to remove this inefficiency.

Finally, we note that while the expectations of traffic increase in cellular networks is by about 92\% per year, the expected data revenue increase is only about 10-18\%. For this reason, reduction of Operational Expenses by improving the energy efficiency in such networks will be extremely important for service providers.
\clearpage
\newpage
\bibliographystyle{mybibstyle}
\small
\bibliography{IEEEabrv,bib/GreenComm}
\end{document}